\documentclass{Rinton-P9x6}
\begin{document}
\title{Remarks on harmonic maps, solitons, and dilaton gravity}
\author{Floyd L. Williams}
\address{
Department of Mathematics, 
University of Massachusetts, \\
Amherst, Massachusetts 01003, U.S.A. \\
E-mail: williams@math.umass.edu}
\maketitle

\abstracts{
Another connection of harmonic maps to gravity is presented.
Using 1-soliton and anti-soliton solutions of the sine-Gordon 
equation, we construct a pair of harmonic maps that we express 
in terms of a particular dilaton field in Jackiw-Teitelboim 
gravity. This field satisfies a linearized sine-Gordon equation.
We use it also to construct an explicit transformation that 
relates the corresponding solitonic metric to a two dimensional 
black hole metric.
}

\section{Introduction}
The theory of harmonic maps provides a pleasant, unifying setting
in which various field equations can be viewed and discussed.
The equations of motion of a Bosonic string, for example, coincide
with the requirement that the map of its world sheet to 
26-dimensional space should be harmonic. The solution of the 
$O(3)$ $\sigma$-model is provided by a harmonic map from 
the unit 2-sphere $S^2$ to itself. Certain Einstein equations 
are given by harmonic maps. A broad overview of the role
of harmonic maps in Yang-Mills theory, general relativity,
and quantum field theory is presented in the inspiring 
papers of C. Misner{\cite{misner}} and 
N. S$\acute{a}$nchez{\cite{sanchez}}, for example.

We consider a pair of harmonic maps from the plane $R^2$
to $S^2$ that we relate to sine-Gordon solitons and
to two dimensional Jackiw-Teitelboim dilaton gravity.
An intriguing observation of J. Gegenberg and
G. Kunstatter connects such solitions
with two dimensional black holes{\cite{gegen1,gegen}}.
We construct an explicit transformation $\Psi$ that
takes the solitonic metric to a black hole metric,
and we express the harmonic maps in terms of the 
dilaton field.

\section{Harmonic maps from $R^2$ to $S^2$}

A smooth map $\Phi: (M,g) \rightarrow (N,h)$
of Riemannian (or pseudo Riemannian) manifolds is
called harmonic if it satisfies the following (local)
conditions; one can also formulate harmonicity by a global
condition{\cite{eells}}.
Let $(U, \phi = (x_1,\ldots,x_m))$
and $(V, \psi = (y_1,\ldots,y_n))$
be coordinate systems on $M$,$N$ with 
$U \subset \Phi^{-1} (V)$,
let $\Phi^j = y_j \circ \Phi \circ \phi^{-1} (1 \leq j \leq n)$
denote the $j^{th}$ component of $\Phi$ relative to these 
systems, let $\Gamma^k_{ij}$ denote the Christoffel symbols
of $(N,h)$, and let
$\partial_i = \frac{\partial}{\partial x_i} (1 \leq i \leq m)$.
If
\begin{equation}
B = 
\frac{1}{\sqrt{\left| \det (g \circ \phi^{-1}) \right|}}
\sum_{i,j=1}^{m}
\partial_i
{\sqrt{\left| \det (g \circ \phi^{-1}) \right|}}
(g^{ij} \circ \phi^{-1})
\partial_j
\end{equation}
is the Laplacian of $(M,g)$ on $\phi(U)$,
then we require that
\begin{equation}
( \tilde{B}_s \Phi ) (p)
\equiv
\sum_{i,j=1}^m
\left. \left. 
\left.
(g^{ij} \circ \phi^{-1})
\sum_{k,r=1}^n \partial_i \Phi^k \partial_j \Phi^r
\right|_{\phi(p)}
\Gamma_{kr}^s (\Phi (p))
+
( B \Phi^s ) \right|_{\phi(p)}
\right.
= 0
\end{equation}
for $p \in U$, $1 \leq s \leq n$.
We construct harmonic maps 
$\Phi = \Phi^{\pm}: R^2 \rightarrow S^2$
as follows.
$\Phi = (\cos \beta \sin \alpha,~ \sin \beta \sin \alpha,~ \cos \alpha)$
where $\alpha,~\beta: R^2 \rightarrow R$
are given by 
$\alpha(x,t) = u(x,t)/2,~~ \beta(x,t) = m (vx +t)/a$
for parameters $m,v > 0, a = a(v) \equiv \sqrt{1 + v^2}$
where for
$\rho (x,t) \equiv m (x - vt)/a,
u(x,t) = u^{\pm}(x,t) = 4 \tan^{-1} e^{\pm \rho (x,t)}$
are one-soliton solutions of the Euclidean sine-Gordon
equation~(SGE)
\begin{equation}
\Delta u 
= \frac{\partial^2 u}{\partial x^2}
+ \frac{\partial^2 u}{\partial t^2}
= m^2 \sin u
.
\label{eqn2.4}
\end{equation}
The harmonicity condition on $\Phi$ in fact reduces
to condition (\ref{eqn2.4}) - which contrasts the point of 
view in \cite{gegen} where (\ref{eqn2.4}) is obtained by
variation of the dilaton field $\tau$ in the 
Jackiw-Teitelboim (J-T) action
\begin{equation}
I(\tau, u)
= \frac{1}{2 G} \int dx \int dt
~\tau \left[ \Delta u - m^2 \sin u \right]
.
\end{equation}
$m$ is revealed as a mass parameter and
$v$ as a soliton velocity parameter. As pointed out in
\cite{gegen} the linearised SGE 
$\Delta \tau = (m^2 \cos u) \tau$
is satisfied by the field 
\begin{equation}
\tau(x,t) = a(v) \mbox{sech} \rho (x,t)
.
\label{eqn2.6}
\end{equation}

\section{Satement of the main result}
The solitonic metric
\begin{equation}
ds^2 
= \cos^2 \alpha(x,t) dx^2
- \sin^2 \alpha(x,t) dt^2
\label{eqn3.1}
\end{equation}
has scalar curvature 
$R = \frac{2 \Delta u}{\sin u}$,
which is therefore constant by (\ref{eqn2.4}): $R = 2 m^2$;
or the Gaussian curvature $K= -\frac{R}{2} = - m^2$.
Recall that $a = a(v) = \sqrt{1 + v^2}$.

\subsection*{Theorem}

Let $\Psi = ( \psi_1, \psi_2 ): R^2 \rightarrow R^2$
be the transformation defined as follows:
\begin{equation}
\begin{array}{rcl}
\psi_1(T,r) 
&=&
vT + \frac{1}{m} \mbox{coth}^{-1}
\left[ \sqrt{a^2 - m^2 r^2} \right]
~,
\\[4pt]
\psi_2(T,r) 
&=&
\frac{\psi_1(T,r)}{v}
- \frac{a}{mv}
\log \left[ \frac{a + \sqrt{a^2 - m^2 r^2}}{m r} \right]
\end{array}
\end{equation} 
on the domain 
\begin{equation}
C^{\dagger} = 
\left\{ \left. (T,r) \in R^2 \right|
0<r< \frac{a}{m}, \sqrt{a^2 - m^2 r^2} > 1 \right\}
,
\end{equation}
and let $\Theta = (\theta_1, \theta_2)): R^2 \rightarrow R^2$
be defined by 
\begin{equation}
\begin{array}{rcl}
\theta_1(x,t)
&=& - \frac{1}{mv} \mbox{coth}^{-1} 
[ a~ \mbox{tanh} \rho(x,t)] + \frac{x}{v}
,\\[4pt]
\theta_2(x,t)
&=& \frac{a}{m} \mbox{sech} \rho(x,t) = \frac{\tau(x,t)}{m}
\end{array}
\end{equation}
on the domain
\begin{equation}
D^{\dagger} 
= \left\{ \left. (x,t) \in R^2 \right|
a~ |\mbox{tanh} \rho (x,t) | > 1 \right\}
. 
\end{equation}
Then $\Psi: C^{\dagger} \rightarrow D^{\dagger}$
and $\Theta: D^{\dagger} \rightarrow C^{\dagger}$
are bijections and inverses of each other:
$\Theta \circ \Psi = 1$
on $C^{\dagger}, \Psi \circ \Theta = 1$ 
on $D^{\dagger}$. Also $\Psi$ transforms the solitonic metric 
(\ref{eqn3.1}) to the black hole metric
\begin{equation}
ds^2 = (M - m^2 r^2) dT^2 - (M - m^2 r^2)^{-1} dr^2
\label{eqn3.6}
\end{equation}
with mass $M = v^2$.
Thus, conversely, $\Theta$ takes (\ref{eqn3.6}) back to
(\ref{eqn3.1}). Also the harmonic maps $\Phi^{\pm}$
constructed in the preceding section are expressed 
in terms of the dilaton field (\ref{eqn2.6}) as follows:
\begin{equation}
\Phi^{\pm}
= \frac{1}{a}
(\tau \cos \beta,~\tau \sin \beta, \pm a~ \mbox{tanh} \rho)
.
\end{equation}

\subsection*{~}

\indent
\hspace*{8mm}
The interesting connection of sine-Gordon solitons
to black hole solitons in J-T gravity is the 
remarkable observation of the paper \cite{gegen},
although the transformation $\Psi$ that we have 
presented here does not explicitely appear there.
We have considered another connection of harmonic
maps to gravity.

\end{document}